\documentclass[a4paper,11pt]{article}
\pdfoutput=1 

\usepackage{jinstpub} 

\usepackage[separate-uncertainty,retain-explicit-plus,per-mode = symbol]{siunitx}
\usepackage{lineno}
\usepackage{listings}
\newcolumntype{d}[1]{D{.}{\cdot}{#1}}
\newcommand{\Figdir}{}
 
\DeclareSIUnit\c{\mbox{$c$}}
\DeclareSIUnit\week{w}
\DeclareSIUnit\year{yr}
\DeclareSIUnit\yr{yr}
\DeclareSIUnit\yr{yr}
\DeclareSIUnit\standard{std}
\DeclareSIUnit\str{sr}
\DeclareSIUnit\ppm{ppm}
\DeclareSIUnit\ppb{ppb}
\DeclareSIUnit\ppt{ppt}
\DeclareSIUnit\pe{PE}
\DeclareSIUnit\spe{SPE}
\DeclareSIUnit\ev{events}
\DeclareSIUnit\hit{hit}
\DeclareSIUnit\hits{hits}
\DeclareSIUnit\bin{(\mbox{5-PE}~bin)}
\DeclareSIUnit\sgm{\mbox{$\sigma$}}
\DeclareSIUnit\rms{RMS}
\DeclareSIUnit\keVr{\mbox{keV$_{\rm nr}$}}
\DeclareSIUnit\keVee{\mbox{keV$_{e{\rm e}}$}}
\DeclareSIUnit\ph{photons}
\DeclareSIUnit\pm{PMT}
\DeclareSIUnit\inch{''}
\DeclareSIUnit\bit{bit}
\DeclareSIUnit\sample{samples}
\DeclareSIUnit\barn{b}
\DeclareSIUnit\bara{bar}
\DeclareSIUnit\Curie{Ci}
\DeclareSIUnit\psi{psi}
\DeclareSIUnit\mK{\milli\kelvin}
\DeclareSIUnit\liveday{\mbox{live-days}}
\DeclareSIUnit\days{\mbox{days}}

\newcommand{\AmBe}{\ensuremath{^{241}}Am\ensuremath{^9}Be}
\newcommand{\AmC}{\ensuremath{^{241}}Am\ensuremath{^{13}}C}

\newcommand{\dsf}{\mbox{DarkSide-50}}

\newcommand{\scene}{\mbox{SCENE}}

\newcommand{\lsv}{\mbox{LSV}}
\newcommand{\wcv}{\mbox{WCV}} 

\newcommand{\tpc}{\mbox{TPC}}

\newcommand{\lar}{\mbox{LAr}}

\newcommand{\DS}{\mbox{DarkSide}}

\newcommand{\SCENE}{\mbox{SCENE}}

\newcommand{\NR}{\mbox{NR}}

\newcommand{\bg}{\mbox{$\beta/\gamma$}}

\newcommand{\SOne}{\mbox{S1}}

\newcommand{\FNinety}{\mbox{f$_{90}$}}
\newcommand{\timefno}{\SI{90}{\nano\second}} 
\newcommand{\fixedintone}{\SI{7}{\micro\second}} 

\newcommand{\lsvscintillatormass}{\SI{30}{\tonne}} 

\newcommand{\brbortenground}{\SI{6.4}{\percent}}
\newcommand{\enbortengroundalpha}{\SI{1775}{\keV}}

\newcommand{\tdrift}{\ensuremath{t_{drift}}}

\title{CALIS - a CALibration Insertion System for the \dsf\, dark matter search experiment}
\newcommand{\APC}{a}
\newcommand{ \AQGSSI}{b}
\newcommand{\AQLNGS}{c}
\newcommand{\Augustana}{d}
\newcommand{\Belgorod}{e}
\newcommand{\Campinas}{f}
\newcommand{\CAINFN}{g}
\newcommand{\CAUni}{h}
\newcommand{\Chicago}{i}
\newcommand{\BHSU}{j}
\newcommand{\Dubna}{k}
\newcommand{\FNAL}{l}
\newcommand{\GEINFN}{m}
\newcommand{\GEUni}{n}
\newcommand{\Hawaii}{o}
\newcommand{\Houston}{p}
\newcommand{\IHEP}{q}
\newcommand{\Kiev}{r}
\newcommand{\Krakow}{s}
\newcommand{\Kurchatov}{t}
\newcommand{\LLNL}{u}
\newcommand{\LPNHE}{v}
\newcommand{\LSC}{w}
\newcommand{\MEPhI}{x}
\newcommand{\MIINFN}{y}
\newcommand{\MIUni}{z}
\newcommand{\MSU}{aa}
\newcommand{\NAINFN}{bb}
\newcommand{\NAUni}{cc}
\newcommand{\PNNL}{dd}
\newcommand{\Petersburg}{ee}
\newcommand{\PGINFN}{ff}
\newcommand{\PGUni}{gg}
\newcommand{\Princeton}{hh}
\newcommand{\RMUnoINFN}{ii}
\newcommand{\RMUnoUni}{jj}
\newcommand{\RMTreINFN}{kk}
\newcommand{\RMTreUni}{ll}
\newcommand{\USP}{mm}
\newcommand{\SLAC}{nn}
\newcommand{\IPHC}{oo}
\newcommand{\Temple}{pp}
\newcommand{\Davis}{qq}
\newcommand{\UCAS}{rr}
\newcommand{\UCLA}{ss}
\newcommand{\UMass}{tt}
\newcommand{\VTech}{uu}

\affiliation[\APC]{APC, Universit\'e Paris Diderot, CNRS/IN2P3, CEA/Irfu, Obs. de Paris, Sorbonne Paris Cit\'e, Paris 75205, France}
\affiliation[\AQGSSI]{Gran Sasso Science Institute, L'Aquila AQ 67100, Italy}
\affiliation[\AQLNGS]{Laboratori Nazionali del Gran Sasso, Assergi AQ 67010, Italy}
\affiliation[\Augustana]{Department of Physics, Augustana University, Sioux Falls, SD 57197, USA}
\affiliation[\Belgorod]{Radiation Physics Laboratory, Belgorod National Research University, Belgorod 308007, Russia}
\affiliation[\Campinas]{Institute of Physics Gleb Wataghin, Universidade Estadual de Campinas, S\~ao Paulo 13083-859, Brazil}
\affiliation[\CAINFN]{Istituto Nazionale di Fisica Nucleare, Sezione di Cagliari, Cagliari 09042, Italy}
\affiliation[\CAUni]{Department of Physics, Universit\`a degli Studi, Cagliari 09042, Italy}
\affiliation[\Chicago]{Kavli Institute, Enrico Fermi Institute, and Dept. of Physics, University of Chicago, Chicago, IL 60637, USA}
\affiliation[\BHSU]{School of Natural Sciences, Black Hills State University, Spearfish, SD 57799, USA}
\affiliation[\Dubna]{Joint Institute for Nuclear Research, Dubna 141980, Russia}
\affiliation[\FNAL]{Fermi National Accelerator Laboratory, Batavia, IL 60510, USA}
\affiliation[\GEINFN]{Istituto Nazionale di Fisica Nucleare, Sezione di Genova, Genova 16146, Italy}
\affiliation[\GEUni]{Department of Physics, Universit\`a degli Studi, Genova 16146, Italy}
\affiliation[\Hawaii]{Department of Physics and Astronomy, University of Hawai'i, Honolulu, HI 96822, USA}
\affiliation[\Houston]{Department of Physics, University of Houston, Houston, TX 77204, USA}
\affiliation[\IHEP]{Institute of High Energy Physics, Beijing 100049, China}
\affiliation[\Kiev]{Institute for Nuclear Research, National Academy of Sciences of Ukraine, Kiev 03680, Ukraine}
\affiliation[\Krakow]{Smoluchowski Institute of Physics, Jagiellonian University, Krakow 30348, Poland}
\affiliation[\Kurchatov]{National Research Centre Kurchatov Institute, Moscow 123182, Russia}
\affiliation[\LLNL]{Lawrence Livermore National Laboratory, Livermore, CA 94550, USA}
\affiliation[\LPNHE]{LPNHE Paris, Universit\'e Pierre et Marie Curie, Universit\'e Paris Diderot, CNRS/IN2P3, Paris 75252, France}
\affiliation[\LSC]{Laboratorio Subterr\'aneo de Canfranc, Canfranc Estaci\'on 22880, Spain}
\affiliation[\MEPhI]{National Research Nuclear University MEPhI, Moscow 115409, Russia}
\affiliation[\MIINFN]{Istituto Nazionale di Fisica Nucleare, Sezione di Milano, Milano 20133, Italy}
\affiliation[\MIUni]{Department of Physics, Universit\`a degli Studi, Milano 20133, Italy}
\affiliation[\MSU]{Skobeltsyn Institute of Nuclear Physics, Lomonosov Moscow State University, Moscow 119991, Russia}
\affiliation[\NAINFN]{Istituto Nazionale di Fisica Nucleare, Sezione di Napoli, Napoli 80126, Italy}
\affiliation[\NAUni]{Department of Physics, Universit\`a degli Studi Federico II, Napoli 80126, Italy}
\affiliation[\PNNL]{Pacific Northwest National Laboratory, Richland, WA 99354, USA}
\affiliation[\Petersburg]{St. Petersburg Nuclear Physics Institute NRC Kurchatov Institute, Gatchina 188350, Russia}
\affiliation[\PGINFN]{Istituto Nazionale di Fisica Nucleare, Sezione di Perugia, Perugia 06123, Italy}
\affiliation[\PGUni]{Department of Chemistry, Biology and Biotechnology, Universit\`a degli Studi, Perugia 06123, Italy}
\affiliation[\Princeton]{Department of Physics, Princeton University, Princeton, NJ 08544, USA}
\affiliation[\RMUnoINFN]{Istituto Nazionale di Fisica Nucleare, Sezione di Roma Uno, Roma 00185, Italy}
\affiliation[\RMUnoUni]{Department of Physics, Universit\`a degli Studi ``La Sapienza''di Roma, Roma 00185, Italy}
\affiliation[\RMTreINFN]{Istituto Nazionale di Fisica Nucleare, Sezione di Roma Tre, Roma 00146, Italy}
\affiliation[\RMTreUni]{Department of Physics and Mathematics, Universit\`a degli Studi Roma Tre, Roma 00146, Italy}
\affiliation[\USP]{Instituto de F\'{i}sica, Universidade de S\~ao Paulo, S\~ao Paulo 05508-090, Brazil}
\affiliation[\SLAC]{SLAC National Accelerator Laboratory, Menlo Park, CA 94025, USA}
\affiliation[\IPHC]{IPHC, Universit\'e de Strasbourg, CNRS/IN2P3, Strasbourg 67037, France}
\affiliation[\Temple]{Department of Physics, Temple University, Philadelphia, PA 19122, USA}
\affiliation[\Davis]{Department of Physics, University of California, Davis, CA 95616, USA}
\affiliation[\UCAS]{School of Physics, University of Chinese Academy of Sciences, Beijing 100049, China}
\affiliation[\UCLA]{Department of Physics and Astronomy, University of California, Los Angeles, CA 90095, USA}
\affiliation[\UMass]{Amherst Center for Fundamental Interactions and Dept. of Physics, University of Massachusetts, Amherst, MA 01003, USA}
\affiliation[\VTech]{Department of Physics, Virginia Tech, Blacksburg, VA 24061, USA}

\author[\APC]{P.~Agnes,}
\author[\USP]{I.~F.~M.~Albuquerque,}
\author[\UMass,\FNAL,\PNNL]{T.~Alexander,}
\author[\Augustana]{A.~K.~Alton,}
\author[\PNNL]{D.~M.~Asner,}
\author[\PNNL]{H.~O.~Back,}
\author[\FNAL]{B.~Baldin,}
\author[\FNAL]{K.~Biery,}
\author[\RMUnoINFN]{V.~Bocci,}
\author[\AQLNGS]{G.~Bonfini,}
% ...
\emailAdd{bernd@hawaii.edu}
\author[\Houston,\UCLA]{A.~L.~Renshaw,}
\author[\RMUnoINFN]{M.~Rescigno,}
\author[\APC]{Q.~Riffard,}
\author[\PGUni,\PGINFN]{A.~Romani,}
\author[\NAINFN,\Princeton]{B.~Rossi,}
\author[\AQLNGS]{N.~Rossi,}
\author[\VTech]{D.~Rountree,}
\author[\AQLNGS]{D.~Sablone,}
\author[\MIINFN]{P.~Saggese,}
\author[\Chicago]{R.~Saldanha,}
\author[\Princeton]{W.~Sands,}
\author[\AQGSSI,\AQLNGS]{C.~Savarese,}
\author[\Davis]{B.~Schlitzer,}
\author[\Campinas]{E.~Segreto,}
\author[\Petersburg]{D.~A.~Semenov,}
\author[\Princeton]{E.~Shields,}
\author[\Houston]{P.~N.~Singh,}
\author[\Kurchatov,\MEPhI]{M.~D.~Skorokhvatov,}
\author[\Dubna]{O.~Smirnov,}
\author[\Dubna]{A.~Sotnikov,}
\author[\Princeton]{C.~Stanford,}
\author[\UCLA,\AQLNGS,\Kurchatov]{Y.~Suvorov,}
\author[\AQLNGS]{R.~Tartaglia,}
\author[\Temple]{J.~Tatarowicz,}
\author[\GEINFN]{G.~Testera,}
\author[\APC]{A.~Tonazzo,}
\author[\NAUni]{P.~Trinchese,}
\author[\Petersburg]{E.~V.~Unzhakov,}
\author[\RMUnoINFN,\RMUnoUni]{M.~Verducci,}
\author[\Dubna]{A.~Vishneva,}
\author[\VTech]{B.~Vogelaar,}
\author[\Princeton]{M.~Wada,}
\author[\NAUni,\NAINFN]{S.~Walker,}
\author[\UCLA]{H.~Wang,}
\author[\IHEP,\UCLA]{Y.~Wang,}
\author[\Temple]{A.~W.~Watson,}
\author[\Princeton]{S.~Westerdale,}
\author[\Temple]{J.~Wilhelmi,}
\author[\Krakow]{M.~M.~Wojcik,}
\author[\Princeton]{Xi.~Xiang,}
\author[\UCLA]{X.~Xiao,}
\author[\Princeton]{J.~Xu,}
\author[\IHEP]{C.~Yang,}
\author[\UMass]{A.~Zec,}
\author[\IHEP]{W.~Zhong,}
\author[\Princeton]{C.~Zhu}
\author[\Krakow]{and G.~Zuzel}
\collaboration{The \DS\ Collaboration} 

\abstract{
This paper describes the design, fabrication, commissioning and use of  a CALibration source Insertion System (CALIS) in the \dsf\ direct dark matter search experiment. CALIS  deploys radioactive sources into the liquid scintillator veto to characterize the detector response and detection efficiency of the \dsf\ Liquid Argon Time Projection Chamber, and the surrounding \lsvscintillatormass\ organic liquid scintillator neutron veto.
It was commissioned in September 2014 and has been used successfully in several gamma and neutron source campaigns since then. A description of the hardware and an excerpt of calibration analysis results are given below.}

\keywords{Dark matter detectors (WIMPs), Noble liquid detectors, Liquid scintillators, Radioactive source calibration, calibration device}

\arxivnumber{1611.02750}
\begin{document}
\maketitle



\section{Introduction}\label{sec:intro}\label{sec:introduction}
\dsf\ is a Liquid Argon Time Projection Chamber (\lar\ \tpc), operated in Italy's Gran Sasso National Laboratory (LNGS) to search for nuclear recoils induced by weakly interacting massive particles (WIMPs). The first physics result was reported in \cite{Agnes:2015gu} based on 50 live data collection days with Atmospheric Argon (AAr). A first WIMP search using argon extracted from underground sources (Underground Argon, UAr) has been reported in \cite{Agnes:2015_uar}, following the WIMP search with AAr. UAr has a lower concentration of the radioactive $\beta$-emitter $^{39}$Ar by a factor (1.4 $\pm$ 0.2) $\times\, 10^3$ relative to AAr. The combination of these two data sets, both background-free, provides the most sensitive limit on a dark matter search using a \lar\ \tpc\ to date with a 90\% CL upper limit on the WIMP-nucleon spin-independent cross section of $2.0 \times 10^{-44}$~cm$^2$ for a WIMP mass of 100 GeV/c$^2$.  
Calibration campaigns have been performed in the presence of AAr and UAr.

\begin{figure}[htbp]
 \centering
\includegraphics[width=\textwidth]{\Figdir 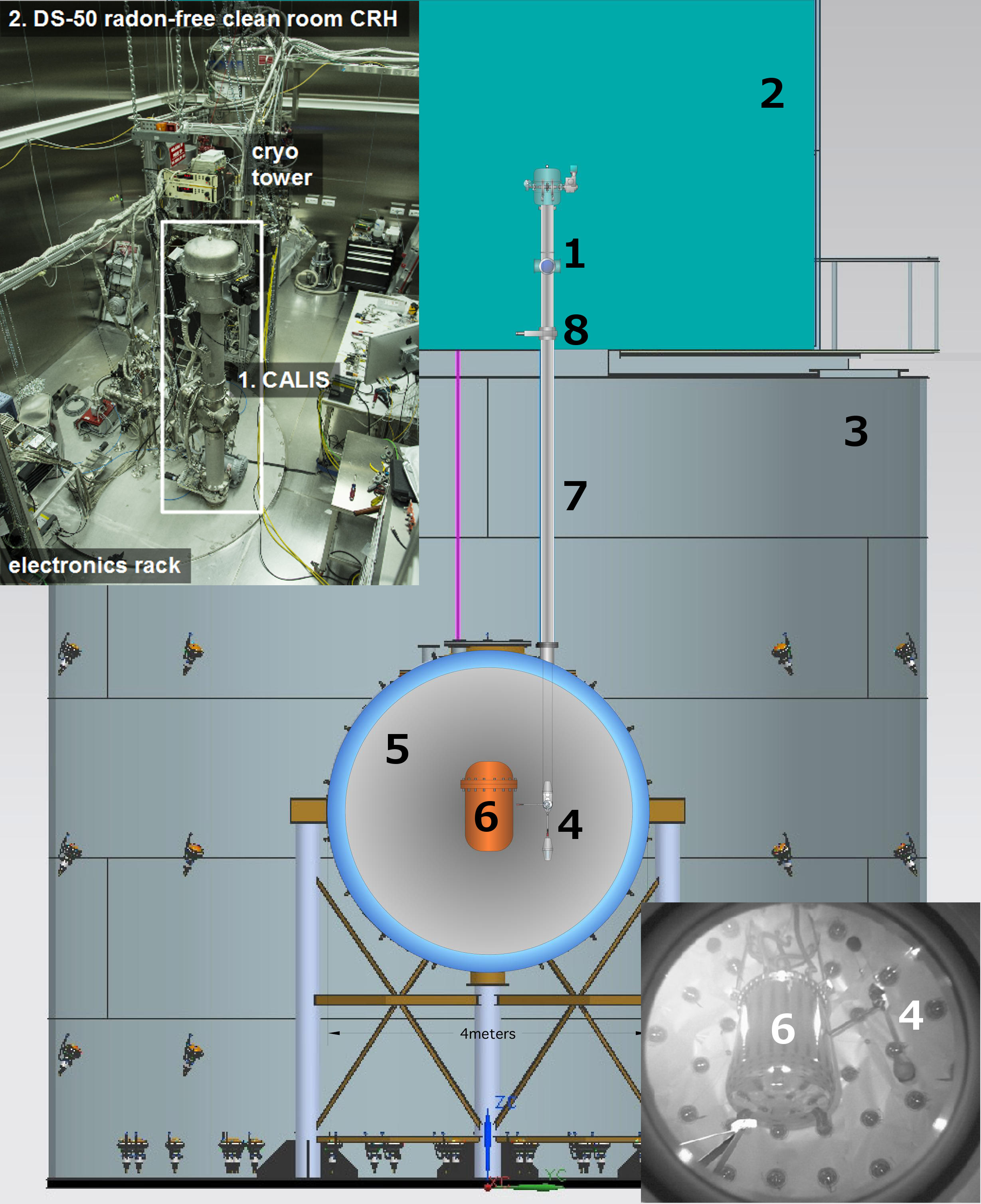}
\caption{A conceptual drawing of CALIS (1) installed in the radon-free clean room, CRH (2), atop the water Cherenkov veto (\wcv, 3). The deployment device (4) which contains the source is deployed in the liquid scintillator veto (LSV, 5) next to the liquid argon time projection chamber's (\lar\ \tpc) cryostat (6). The clean room and the LSV are connected through four access ports, called organ pipes (only one of which is drawn in the above sketch (7)). All four organ pipes end in CRH at gate valves (8) which can be manually opened or closed. During normal operations all four organ pipes are closed. Not included in the drawing are tubes connecting the cryogenic systems in CRH to the cryostat in the \lsv. The top left inset shows CALIS after successful installation inside CRH. The bottom right inset shows a photograph taken with a camera looking upwards into the \lsv\ from the bottom. The deployment device's source arm is articulated and the source in its tip is next to the \lar\ \tpc's cryostat. (The numbering in the insets matches the drawing's.) \label{fig:wholeAssembly_insideDetectors}\label{fig:DS50_with_CALIS}}
\end{figure}

The \dsf\ apparatus is described in detail in \cite{Agnes:2015gu}. As shown in fig.~\ref{fig:wholeAssembly_insideDetectors}, it features a \lar\ \tpc\ surrounded by a 30 t liquid scintillator-based veto (LSV) system, placed inside a water Cherenkov veto detector (\wcv), both of which measure in-situ and suppress radiogenic and cosmogenic backgrounds \cite{Agnes:2015qyz}. On the top of the \wcv\ is a radon-free clean room (CRH for Clean Room Hanoi) housing the cryogenic supply system and electronics (Fig.~\ref{fig:DS50_with_CALIS}). The \lsv\ may be accessed through one of four gate valves in CRH, which are approximately 6 m above the center of the LSV. Each gate valve connects to an access port, called organ pipe, 15 cm in diameter which leads through the WCV and opens into the LSV, 80 cm off the TPC's vertical z-axis.


\section{Design Requirements \& Hardware Implementation} \label{sec:hardware}\label{sec:design_requirements}

\subsection{Purpose \& Deployment Procedure}

Via CALIS radioactive gamma and neutron sources can be deployed inside the LSV to study and calibrate the \tpc\ and the \lsv\ detector response and neutron detection efficiency. For \tpc\ calibration the radioactive source has to be positioned in immediate contact with the cryostat, in order to minimize rate losses through absorption. This is especially important for low energy sources such as $^{57}$Co (122 keV). 

\begin{figure}[htbp]
 \centering
\includegraphics[width=\textwidth]{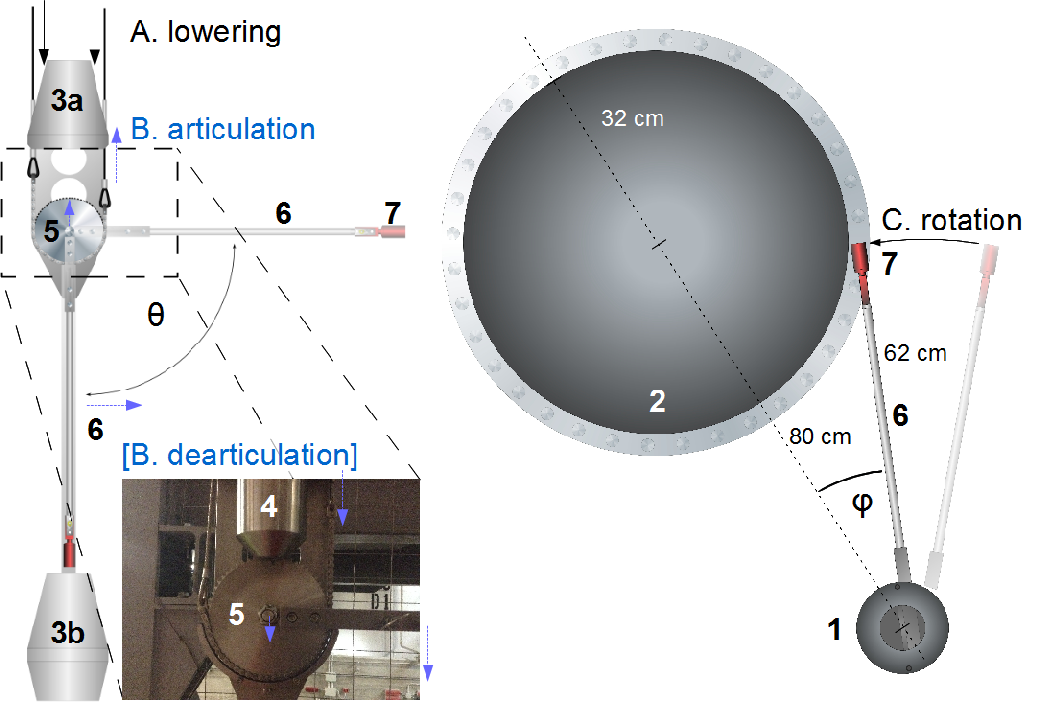} 
  \caption{A side view \textit{(left)} and top view \textit{(right)} of the deployment device (1) next to the \tpc's cryostat (2) is shown. The deployment device features cones at the top and bottom (3a,b), weights (4), an articulation gear (5) and the source arm (6) with the source holder (7) at its tip. A deployment into the \lsv\ involves three steps: \textit{A. lowering} the device into the \lsv\ with the source arm dearticulated, \textcolor{blue}{\textit{B. articulation} of the source arm to horizontal, while the arm is still rotated away in the $xy$-plane and} \textit{C. rotation} of the device to have the source in contact with the cryostat. (In order to distinguish the '\textcolor{blue}{articulation}' from the 'lowering', arrows and captions related to articulation are highlighted in blue in the figure.) Articulation can be set to any angle $\theta$ between 0$^{\circ}$ and 90$^{\circ}$, with horizontal articulation corresponding to 90$^{\circ}$ and dearticulation to 0$^{\circ}$. During calibration campaigns by default the arm length was 62 cm and the source arm was articulated to horizontal. The cryostat's outer radius is 32 cm, the organ pipe is off from the TPC's vertical z-axis by 80 cm.\label{fig:DeploymentDevice}}
\end{figure}

As shown in Fig.~\ref{fig:DeploymentDevice}, right, there is a gap between the 32 cm wide cryostat and the organ pipe, whose central axis is 80 cm away from the TPC's vertical z-axis. This gap precludes a single cable solution deployed from within a glove box as used in several other scintillator experiments \cite{Banks:2014hra, Huang:2013uxa}. 
Instead our apparatus consists of an enclosure which has been installed in CRH and the deployment device featuring an articulation mechanism. Inside the enclosure the deployment device is mounted through two stainless steel cables wound around cable spools. The deployment device contains the mechanical support structure for the source arm, which holds the calibration source at its tip and the gear to articulate it. Focusing just on the mechanics, i.e.~after successful source insertion, a deployment consists of three steps as illustrated in Fig.~\ref{fig:DeploymentDevice}:
\begin{enumerate}
\renewcommand{\theenumi}{\Alph{enumi}} 
\item Lowering the deployment device into the \lsv\ with the source arm dearticulated,
\item articulation of the source arm to horizontal, while the arm is still rotated away from the cryostat in the $xy$-plane and
\item the rotation of the device in the $xy$-plane to bring the calibation source into contact with the cryostat.
\end{enumerate}
In order to return the deployment device from the \lsv\ the sequence and every step is inversed.

\subsection{Deployment \& Articulation Mechanism}\label{sec:DeploymentArticulation}

\begin{figure}[htbp]
 \centering
\includegraphics[width=0.8\textwidth]{\Figdir 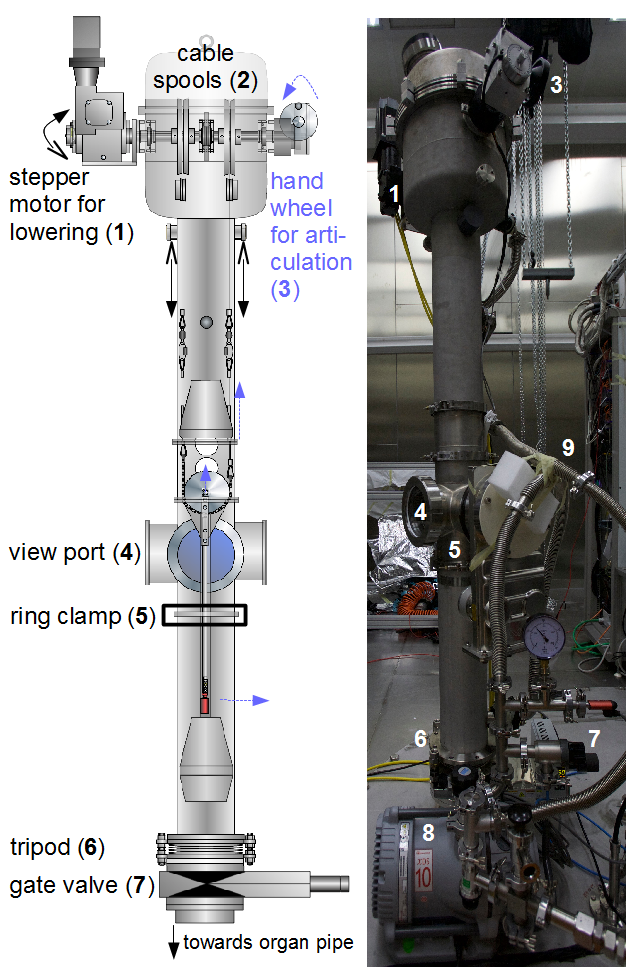}
 \caption{Mechanical drawing of CALIS showing the housing and the deployment device in its home position next to a photo of CALIS as installed inside the clean room CRH. The total height is approx.~240 cm including the gate valve. The 'lowering' and 'articulation' modes of operation are illustrated, the latter is highlighted in blue to distinguish it from the former: In order to move the deployment device down into the \lsv\ (or back up), the stepper motor moves both cable spools (2) concurrently. \textcolor{blue}{In order to articulate, the articulation wheel (3) is rotated manually, which affects only the right spool, thereby shortening the right cable with respect to the left cable, engaging the gear, articulating the arm and lifting the pivot center.} The amount of lifting and the amount of rotations until a horizontal articulation is reached has been calibrated prior to installation in CRH (Sec.~\ref{sec:Testing}). Through the view port (4) the source arm is manipulated. The ring clamp (5) is used for rotations in the $xy$-plane and the tripod (6) has been used for vertical alignment of CALIS wrt.~the organ pipe, which is closed by the gate valve (7). 
In the photograph the vacuum pump (8) and tubing (9) are shown, which are part of the evacuation and purging system (Sec.~\ref{sec:EvacPurge}). \label{fig:CALISDimensions}\label{fig:CALISMechanism}\label{fig:gearDrawing}\label{fig:flushing_purging}
}
\end{figure}

\begin{figure}[htbp]
 \centering
 \includegraphics[width=0.8\textwidth]{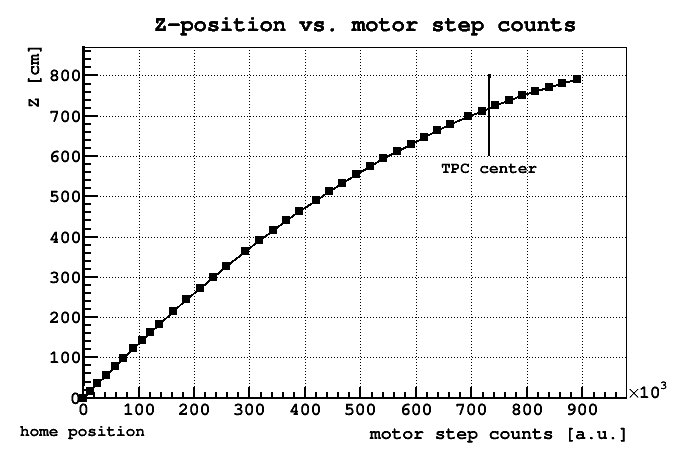}
 \caption{Plot of the deployment device's $z$-position versus motor step counts. The non-linear correspondence between number of steps and cable length deployed arises as follows: As the cables wind around their spools, the winding radius changes, increasing as the deployment device is lifted and decreasing as it is lowered. A motor step count corresponds to a fixed angular distance \textit{d$\theta$}, yet the amount of cable deployed during this motor step is \textit{winding radius $\cdot$ d$\theta$}. As the winding radius changes as a function of $z$-position, the fraction of deployed cable per motor step count changes. The home position is at 0, the TPC center is reached after more than 7 m of travel into the \lsv, the maximum depth is reached at nearly 8 m.}
 \label{fig:z_test}
\end{figure}

As shown in Fig.~\ref{fig:CALISMechanism}, the stepper motor moves both cable spools concurrently and thereby sends the deployment device into the \lsv. An absolute encoder provides its current position, even in the absence of power. The stepper motor is controlled via a simple graphical LabVIEW interface, run on a dedicated laptop, in which the current $z$-position is shown and a target $z$-position can be provided by the operator. $Z$-positions are given in motor step counts, an arbitrary unit which has been calibrated outside CRH in meters (Fig.~\ref{fig:z_test}), as well as relative to the TPC using \tdrift\ distributions from calibration data (Sec.~\ref{sec:SourcePosition}). \label{sec:Nonlinearity:MotorStepCounts}
The non-linearity between motor steps and cable length shown in Fig.~\ref{fig:z_test} arises as follows: As the cables wind around their spools, the winding radius \textit{r$_w$} changes, increasing as the deployment device is lifted and decreasing as it is lowered. A motor step count corresponds to a fixed angular distance \textit{d$\theta$}, yet the amount of cable deployed during this motor step is \textit{$r_w\,\cdot\,d\theta$}. As the winding radius changes as a function of $z$-position, the fraction of deployed cable per motor step count changes.

Source arm articulation is done manually via the articulation wheel. This affects only the cable spool close to the articulation wheel, the right one in Fig.~\ref{fig:CALISMechanism}, thereby shortening the right cable with respect to the left cable and engaging the gear through a chain. As a result the source arm is articulated and the pivot center is lifted. The non-linearity between motor steps and cable length affects also the amount of rotation required by the hand wheel for horizontal source arm articulation. Degrees on the articulation wheel corresponding to a horizontal articulation have been calibrated as a function of $z$-position prior to installation in CRH.

Articulation and a movement in $z$-direction are mutually exclusive since arm articulation leads to more cable wound up on the spool close to the articulation wheel with respect to the other. If then, in $z$-movement mode, both spools would be rotated simultaneously with the same angular speed, the cable close to the articulation wheel would wind up faster than the other, leading to a build up of difference in cable length and the deployment device would only be hanging on one cable. In order to avoid an imbalanced $z$-movement, the arm has to be dearticulated fully before a change in $z$-position can be initiated. This is enforced by an electric switch preventing $z$-movement, which is disengaged only when the source arm is fully dearticulated (i.e.~hanging vertically). 

\subsubsection*{Deployment Device}
As shown in Fig.~\ref{fig:DeploymentDevice} the device is equipped with tapered cones on its top and bottom, ensuring that the ends do not get snagged on inner edges as it moves down and up, such as when reentering the organ pipe from the inside of the \lsv\ while on its way back to its home position inside the CALIS enclosure. It is attached to the housing by two cables. Swivel hooks are employed in the attachment of the cables to the deployment device allowing them to move freely and not get tangled. 
There are two weights built into the device, one cylindrical in the conical cap above the rotation gear mechanism and one inside the cone at the device's bottom end. Both help to minimize any lateral motion or oscillations during deployment, articulation and dearticulation. They also ensure smooth motion of the deployment device into the organ pipe and back to the home position.


\subsection{CALIS Enclosure \& Scintillator}

Besides providing mechanical support for the deployment device via the cable spools, the CALIS enclosure is an important interface between the radon-free clean room CRH and the \lsv, through which sources are exchanged. 
The enclosure protects the liquid scintillator (LS) and prevents human contact with any traces of harmful LS vapor (Fig.~\ref{fig:CALISMechanism}). It plays the same role as a glove box for similar calibration systems, yet with a narrower foot print inside CRH. The liquid scintillator is a mixture of pseudocumene (PC) and trimethyl borate (TMB) with the wavelength shifter 2,5-diphenyloxazole (PPO) \cite{Agnes:2015qyz}.\footnote{The concentration of TMB and PPO have been varied during campaigns (see Sec.~\ref{sec:CalibCampaigns}).} 
It must not be exposed to oxygen or water that is present in normal clean room air. Contamination of the LS with $^{222}$Rn and its long-lived radioactive daughters has to be avoided, as well. 

Going up from the gate valve on which CALIS has been installed, there is a teflon disk that can electrically isolate CALIS from ground, even though during normal operations the CALIS housing is connected to ground. A tripod with a bellow has been used to vertically align the enclosure right after installation on the gate valve. The bellow is connected to a 59.4 cm long cylindrical stainless steel enclosure pipe. It has the same diameter as the organ pipe (15 cm) and is connected to the view port by a ring clamp, which plays a critical role for rotations in the $xy$-plane (see Sec.~\ref{sec:XYrotation}). The view port can be opened for handling the source arm and exchanging calibration sources. Everything above the ring clamp forms the upper assembly. It features a stainless steel cylindrical enclosure housing the cable drive mechanism, including the cable spools, the stepper motor and the articulation mechanism already described in Sec.~\ref{sec:DeploymentArticulation}. 

\subsubsection*{Vacuum evacuation (flushing) and nitrogen purging system}\label{sec:EvacPurge}
One of this system's most important safety features is making sure that TMB and PC residue on the deployment device are extracted from CALIS and vented prior to opening the view port to access the source arm. This is important for safe working conditions inside CRH as well as for the scintillator and its radiopurity. 

After the source arm insertion and view port closure, the inside of the CALIS housing is filled with normal air that is damaging to the scintillator. A sequence of evacuations and nitrogen purges reduces the fraction of normal air and its contaminants in the air-N$_2$ mixture to negligible levels. Only after this sequence is finalized, the gate valve is opened and the deployment device is introduced into the \lsv. Evacuation is achieved with a vacuum pump and the exhaust air is removed through dedicated vent lines (Fig.~\ref{fig:flushing_purging}).

At the end of a calibration campaign, after the deployment device has returned to its home position inside the enclosure and the gate valve has been closed, scintillator vapor, and in particular TMB, has to be removed prior to opening the view port. Again a sequence of evacuation and nitrogen purges is employed. By lowering the pressure inside of CALIS below the TMB and PC vapor pressure, the scintillator evaporates and is removed through the vacuum pump vent line. Once pressure inside the housing stays consistently below the vapor pressure of TMB, all scintillator has been removed and the view port can be opened to access the source arm.
 
\subsubsection*{Material Compatibility}
All materials coming in contact with the scintillator veto are made of stainless steel or teflon, except for the sealing o-rings which are made of viton. All three materials are certified for contact with all scintillator components including TMB and PC.

\subsection{Source Arms and Source Holder}
\begin{figure}[htbp]
 \centering
  \includegraphics[width=0.7\textwidth]{\Figdir 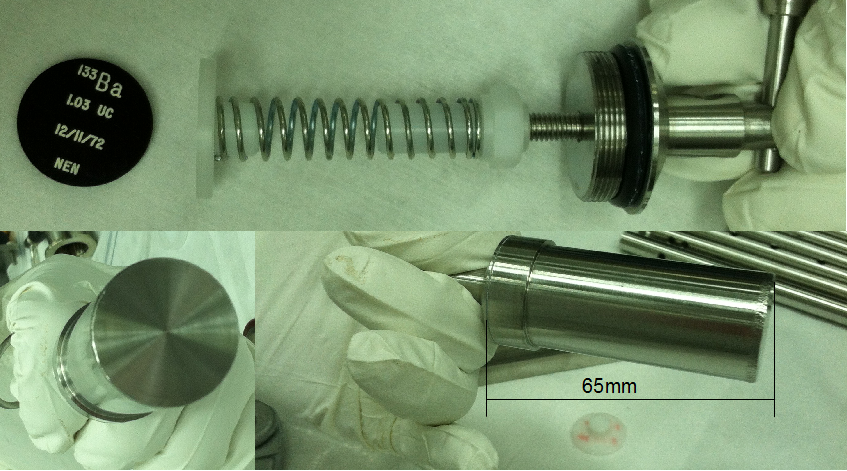}
  \caption{The source holder that connects to an arm and to the articulation gear of the deployment device. The source, here a $^{133}$Ba source, is pressed to the tip of the source holder via a spring.}
  \label{fig:SourceHolder}
\end{figure}

A source arm and source holder are attached to an articulation gear. Different arm lengths have been prepared, with a maximum arm length of 62 cm as measured from the pivot point of the rotation gear to the source holder's tip. The radioactive source is housed inside the source holder (Fig.~\ref{fig:SourceHolder}). During deployment it is pressed to the tip and held in place via a spring. The source holder is sealed such that no liquid scintillator can enter during deployment. This has also been verified during each source extraction and no liquid traces have been found inside.

\subsection{Hardware Details and Safety Features}\label{sec:HardwareDetails}\label{sec:SafetyFeatures}
CALIS offers various safety features to ensure that this device runs smoothly, that no components are lost inside the detector, that any detector contamination by dirty or incompatible materials is avoided, that pressure is maintained and that the introduction of oxygen or water into the LS and TMB is avoided. 

\begin{description}

\item[Cable strength:]
The cables holding the deployment device are rated for loads over 600\,kg, while the device's weight is at the level of 10-15\,kg, thus well below the cables' breaking strength.

\item[Drive mechanism:]
A magnetic brake, which releases only in the presence of electric power, ensures that the deployment device does not move, if a power failure occurs. Servo motor torque is limited in case of an unexpected load and the risk of breaking the cable is avoided.

The speed reducer is implemented in a double worm gear design. The primary worm gear has a 50:1 reduction and the secondary worm has an 82:1 reduction. The servo motor input speed is 2400 RPMs, the output is 0.6 RPM and its weight capacity is 67\,kg. If a power failure occurs, the speed reducer can hold the load at any position. The motor speed has been limited to 0.4\,cm/s, minimizing any lateral oscillation of the deployment device during lowering and raising of the source. This is also the maximum speed at which the motor does not overheat.

\item[Light and leak tightness of CALIS:]
When data is taken with the \lsv\ while the gate valve is open, as is the case during calibration campaigns with CALIS, absolute light tightness and pressure leak tightness is required. All view ports are covered with light tight covers when the gate valve is open. Both light and leak tightness was extensively validated throughout the manufacturing process, commissioning and during calibration campaigns.

\item[Securing source:] 
All connection points for the source and the source arm have been secured with two push locking pins that cannot be disengaged without a person pressing the pin. In addition, the source holder and its two locking pins are all tethered from outside the view port until they are locked in place, preventing them from accidentally falling into the interior of the CALIS housing.

\item[Manual retraction system:]
It is possible to manually retract the deployment device back to its home position and to close the gate valve in the unlikely event of a complete motor failure while the deployment device is deployed. The motor is disengaged, and a wrench is used to manually wind the cables back on the spools and to retract the deployment device above the gate valve. During this procedure, the nitrogen blanket protecting the \lsv\ is preserved. 
   
\item[High limit switch:]
A high limit switch is a hardware interlock that prevents the deployment device from hitting cable spools and gears, should it pass beyond the home position in the CALIS housing (Fig.~\ref{fig:CALISMechanism}). 
\end{description}

Thus far during calibration campaigns it has not been necessary to use the manual retraction system nor has the high limit switch been activated.
	

\subsection{Degrees of Freedom}

CALIS is capable of deploying sources at various positions inside the \lsv. Besides movement along $z$ down to its maximum cable length, it is possible to articulate at an angle $\theta$ between 0$^{\circ}$ and 90$^{\circ}$, where $\theta$ is the zenith-angle (Fig.~\ref{fig:DeploymentDevice}). Angles of more than 90$^{\circ}$ are excluded because the articulation chain's end is reached at a 90$^{\circ}$ angle (see photo inset in Fig.~\ref{fig:DeploymentDevice}).

\subsubsection*{$XY$-plane rotation}\label{sec:XYrotation}
The connection below the view port has an o-ring seal and uses a ring clamp to compress the seal. This clamp can be slightly loosened allowing the upper assembly (everything above and including the view port) to be rotated with respect to the lower assembly and the \tpc. Rotation in the $xy$-plane can even be performed while the device is deployed next to the cryostat, since the seal is helium leak and light tight even when loosened.

In principle a rotation by an arbitrary angle can be done, except when the arm would interfere with the cryostat. In one calibration campaign an \AmBe\ neutron source was deployed directly next to the cryostat and rotated away to an angle $\phi\,=\,90^\circ$.  A relative comparison between the two data sets as well as comparisons with Monte Carlo simulations gave insights on optical shadowing effects by the cryostat (Sec.~\ref{sec:CalibCampaigns}). 


\begin{figure}[htbp]
 \centering
  \includegraphics[width=0.7\textwidth]{\Figdir 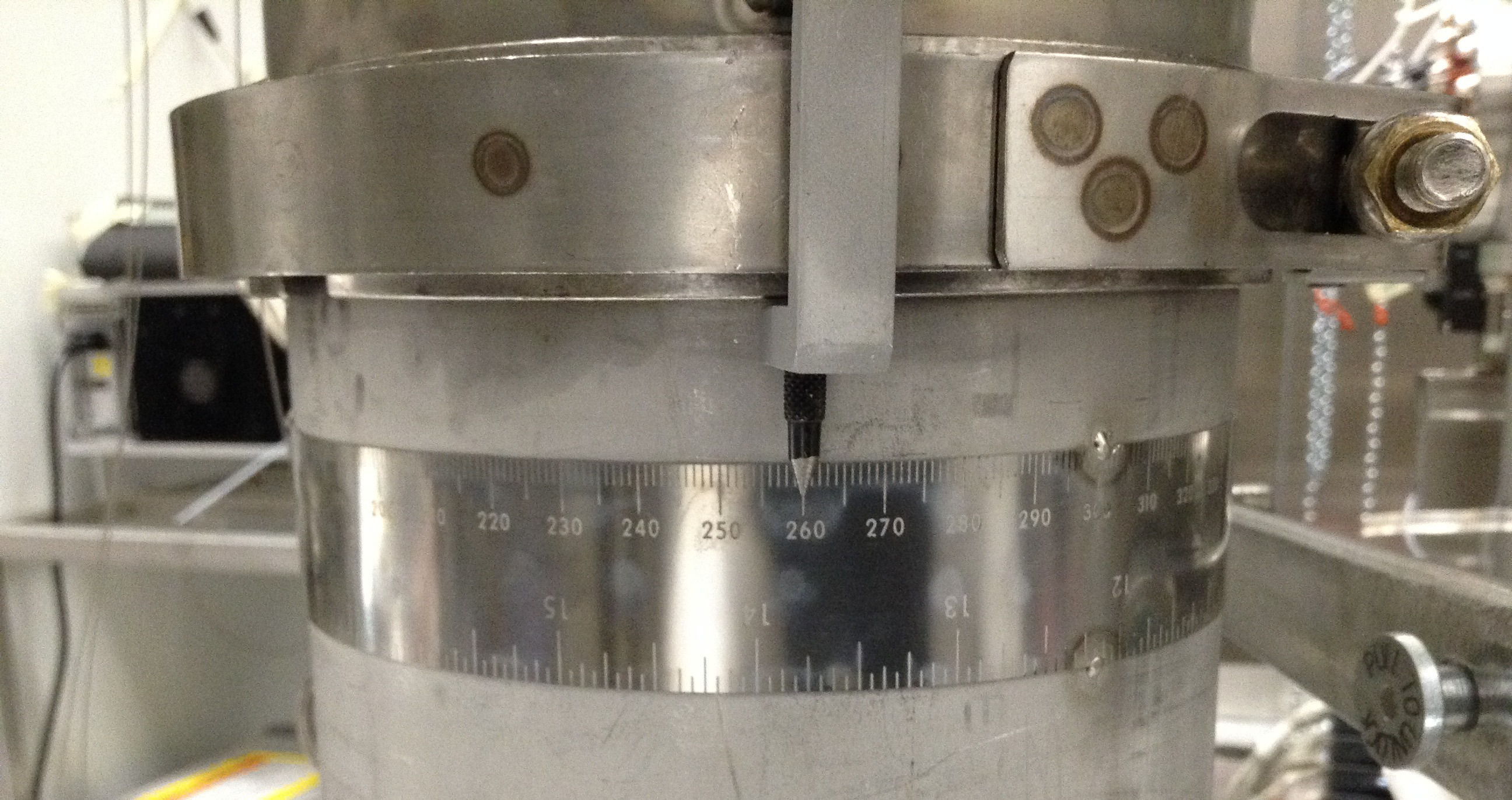}
  \caption{Beneath the view port is a ring clamp with a ruler underneath. The rotation angle is read from the ruler going around the pipe. The ruler is in mm, which has been calibrated in degrees. To perform azimuthal rotation, the ring clamp is slightly loosened, and the entire upper assembly is rotated with respect to the lower assembly, along with the deployment device.}
  \label{fig:ring_clamp}
\end{figure} 

\subsubsection*{``No fly'' zone}
A ``no fly'' zone is defined directly above the cryostat where there are many TPC supply tubes. The source arm may not enter in this region, hence CALIS operators, who are specifically trained personnel to manipulate CALIS, are required to avoid articulation in this region. Since articulation is done manually and slowly by the operator, accidental entanglement of the source arm in the supply tubes is extremely unlikely.


\subsubsection*{Default configuration}\label{sec:DefaultConfig}\label{sec:CentralPosition}
By default, the deployment device has been deployed with the longest source arm (62 cm) in a horizontal position in contact with the cryostat and to our 'central position' in $z$. The 'central position' is a motor step position, which we calibrated using $t_{drift}$ distributions of $^{133}$Ba to be within two centimeter of the TPC's active volume center (see Fig.~\ref{fig:SourcePosition}). 

Other degrees of freedom could involve shorter arm lengths, while longer arm lengths would require hardware modifications on the deployment device. Out of four organ pipes, a second one is available for source calibration. (Two organ pipes are not available due to interference with existing infrastructure: the cryogenic tower and the electronics rack.) Moving CALIS to a different organ pipe requires a partial disassembly of CALIS and reinstallation on the other organ pipe's gate valve.

\section{Testing, Cleaning and Commissioning} \label{sec:Testing}\label{sec:Commissioning}
CALIS was assembled at Fermi National Accelerator Laboratory (FNAL) from components produced at FNAL and at the University of Hawaii. Initial basic functionality tests at FNAL included deploying the system in the high bay and making positioning precision measurements over its entire length. CALIS was positioned on a high platform and a laser ranger positioned on the floor was used to measure the distance between the floor and piece of white paper, attached to the bottom of the deployment device, and used as a target for the laser ranger. The system went through more than 30 consecutive cycles, to confirm positioning precision, stability of the system, lack of slippage and lack of uncertainty accrual.  The table relating the stepper motor count and deployment level was generated based on these measurements. This is important since the cable's single wind corresponds to different deployment lengths at different heights, which is the feature of the design. As a result of these measurements, the z positioning precision was confirmed to be 2\,mm, consistent with the unevenness of the paper target in the test setup. CALIS is deployed at low speed of 4\,mm/s to avoid lateral motion during deployment. Deployment to its lowest position confirmed lack of lateral motion during vertical deployment. Swing was also measured during the articulation and de-articulation of the source arm, used to bring the source close to the cryostat. Articulation and de-articulation of the source arm close to a mock-up cryostat was recorded on the video, and lateral motion was analyzed using the video. 1.5\,cm swing was measured and it took about 2\,min for the system to come to rest in air. During deployment in the detector, this swing is damped by the surrounding oil. Lateral motion was also checked during azimuthal motion of the articulated source arm, and no swinging has been observed. Repeatibility of the articulation of the source arm was tested as well as the motion of the deployment device upwards as a result of articulation. The upward motion was measured along with offsets during articulation and de-articulation process. While the source arm articulated to 90$^\circ$ every time, de-articulation resulted in 2$^\circ$ offset from vertical. Further investigation indicated that any accumulation of tension in the cables may result in this offset. As a result, the deployment device is always deployed to the lowest position first, in which the cables are fully unwound, prior to positioning at the desired height, to release any accumulated tension in the cables. CALIS needs to be leak tight, both when the clamp for azimuthal rotation is secure and when it is loose. The leak tightness of CALIS was veriﬁed with helium leak testing during deployment and azimuthal rotation and shown to be leaktight. The limit switch, which prevents the deployment device from moving in a vertical direction while the arm is articulated was found to work consistently. Additionally, an upper limit switch which stops the deployment device from moving too far into the upper assembly (and possibly colliding with the cable spools) was also found to work consistently. Another safety feature of CALIS are the safety clips. These are attached to the outside ofthe lower assembly and connected to each other by stainless steel wire. During the change of the source all locking pins are hooked to safety clips to eliminate any possibilty of dropping parts on the bottom of the lower assembly.  CALIS was shipped pre-assembled to LNGS, where it underwent a comprehensive testing and calibration program along the same lines as the testing done at Fermilab to verify its good operational state as well as the good working condition of its safety limit switches. While still outside the clean room CRH, CALIS was installed on a high bay platform in the LNGS underground Hall C and tested to its full deployment length. Detailed mapping between the deployment location and stepper motor count was aquired. Besides testing basic $z$-motion, articulation and $xy$-rotation, all details of CALIS operations were tested. These included the testing of the functionality of motor controls, the high limit and articulation switch. Validation was concluded with recovery scenarios after, e.g.,~power failures during deployment and motor malfunction during deployment. To prevent any catastrophic failures in these cases, the system was equipped with an absolute encoder, to record absolute stepper motor count even in the case of power failure. The torque of the servo motor is limited in case of an unexpected load which was also tested. Iin the case of motor failure, the stepper motor is disengaged and the deployment device is retrieved above the gate valve by manually winding the cables on the spool with a wrench, and the procedure was verified prior to installation.

An important aspect was the $z$-position calibration of the source as a function of cable length before and after source arm articulation, which is a nonlinear function of motor step counts, as mentioned in Sec.~\ref{sec:Nonlinearity:MotorStepCounts}. Furthermore, the source's $xy$- and $z$-position accuracy and precision were estimated.

This testing campaign's results were reviewed by an internal review board and approval for installation of CALIS inside CRH was granted subsequently. CALIS was cleaned according to official cleaning procedures and installed on its gate valve in September 2014.
After installation on the gate valve a testing focus was the system's light and helium leak tightness, as well as validation of nitrogen and vacuum systems, as these could only be tested fully after installation. A more detailed description of tests performed at FNAL and LNGS can be found in \cite{thesis:Hackett, thesis:Edkins}.

\subsection*{$XY$- and $Z$-position}
Tests in air and in the LSV's scintillator revealed that the dominant source of uncertainty in the source position arises during articulation. Positioning in $z$ before articulation is highly accurate and precise: deployment speed is very low, barely visible to the naked eye (0.4 cm/s), which minimizes lateral motion and avoids contact with the housing or the organ pipe during deployment. Yet during articulation a swing of the source in $xy$ arises from tiny laterally imbalanced forces originating in the cable manipulated for articulation. 

To ensure deployment precision, a procedure has been worked out to make reliably gentle contact with the cryostat, thereby eliminating precision uncertainty in $xy$: After positioning the deployment device in $z$, the source arm is articulated to horizontal while it is still pointing away from the cryostat. Only then is the source brought into contact with the cryostat through an $xy$-rotation of the upper assembly, while monitoring count rates in the TPC's photomultiplier tubes (PMT).  These increase while the source is approaching, yet plateau as soon as contact with the cryostat is made, even if $xy$-rotation continues. This procedure provides a reliable $xy$- and $z$-position for the calibration source and was used throughout all calibration campaigns.


\section{Calibration Campaigns}\label{sec:CalibCampaigns}
\subsection{Laser Calibration}
\wcv, \lsv\ and \tpc\ electronics gain calibration is performed with dedicated laser systems in each of the three subdetectors. During laser runs digitizer pulse integral spectra are recorded at an average occupancy of $< 0.1$ photo electrons (PE) for most PMTs. For each PMT  the mean and variance of the recorded single PE spectrum is estimated. Laser calibration runs are performed periodically during regular dark matter data taking, which means at least daily in case of the \tpc\ \cite{Agnes:2015gu}. 

These laser runs are also an integral part of all calibration campaigns. A laser run is taken at least at the same frequency as during regular data taking and additionally after each change in DAQ, \tpc\ or CALIS configuration, such as drift field changes or source position changes. \wcv\ laser calibration is sufficient to veto muons (and their secondaries) with high efficiency and no source deployments in the \wcv\ have been performed.

\subsection{Radioactive Sources}
\subsubsection{Internal $^{39}$Ar and $^{83m}$Kr Calibration Campaigns}
The \tpc\ detector response, in particular the light yield, has been calibrated using the internal $^{39}$Ar inherent to AAr, as well as $^{83m}$Kr, that has been added into the \lar\ recirculation system during dedicated calibration campaigns \cite{Agnes:2015gu}.

\subsubsection{Gamma Sources}
\begin{figure}[htbp]
 \centering
 \includegraphics[width=0.73\textwidth]{\Figdir 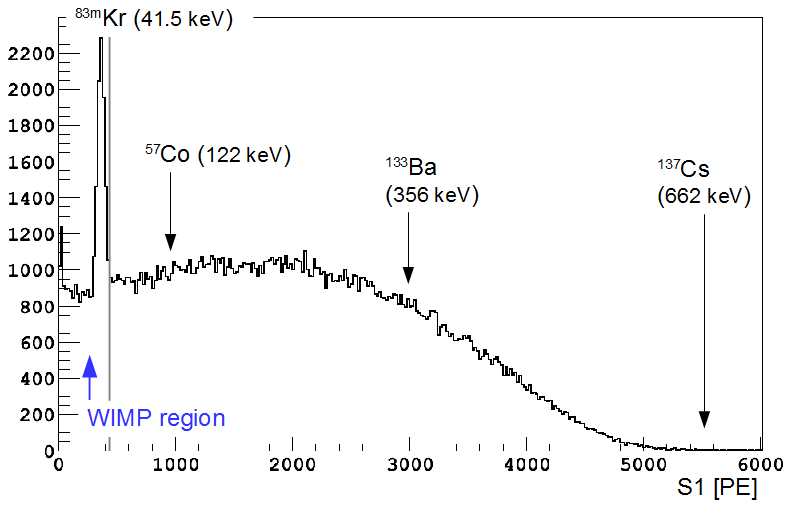}
 \caption{The AAr scintillation spectrum (S1) at null field showing a $^{83m}$Kr peak on top of the internal $^{39}$Ar $\beta$-spectrum inherent to AAr. The positions of full absorption peaks of three gamma sources are indicated and cover the full range of the $^{39}$Ar spectrum.
\label{fig:GammaSources_Ar39spectrum}}
\end{figure}

For the calibration of the \lsv\ detector response and the \tpc's response to electron recoils (ER), $^{57}$Co, $^{133}$Ba and $^{137}$Cs were selected. $^{22}$Na was deployed in a later calibration campaign. They allow a confirmation of PE-energy calibrations from $^{83m}$Kr and the internal $^{39}$Ar, as they cover the $^{39}$Ar energy range (see also Table~\ref{tbl:GammaSources} and Fig.~\ref{fig:GammaSources_Ar39spectrum}). 

After a preselection of gamma source energies, detailed studies with the DarkSide Monte Carlo (MC) simulation package G4DS \cite{DS50:G4DS:paper} were performed to select appropriate source activities and to check the feasibility and physics reach of various deployment positions. Sources with suitable activities were identified for deployment considering also constraints from the \lsv\ and \tpc\ DAQs (Table~\ref{tbl:GammaSources}).

\begin{table}[htbp]
\centering
\caption{The internal $^{39}$Ar inherent to AAr, the $^{83m}$Kr calibration source and gamma sources deployed in DS-50. The $^{83m}$Kr nuclei are produced in decays of $^{83}$Rb which has a half-life of 86.2 days. The $^{83m}$Kr decays with a half-life of 1.8 hours, emitting 32.1 keV and 9.4 keV conversion electrons \cite{Lippincott:2010jb}. The $^{83m}$Kr source activity varied from campaign to campaign, due to the relatively short half-life of $^{83}$Rb, yet it was in the range of a few Bq to some tens of Bq. The $^{39}$Ar activity has been approximately 50 Bq (1 Bq/kg) during AAr filling and negligible in the UAr phase.}
\centering
\begin{tabular}{|l|l|l|l|l|}
\hline
\textbf{source} & \textbf{type} & \textbf{energy} & \textbf{half life} & \textbf{activity} \\ \hline
$^{39}$Ar & $\beta$ &  565 keV endpoint& 269 y  & 50 Bq\\ \hline
$^{83m}$Kr & 2 $\beta$ &  32.1 keV + 9.4 keV & 86.2 d & varying\\ \hline\hline
$^{57}$Co & $\gamma$ & 122 keV & 0.744 y  & 35 kBq \\ \hline
$^{133}$Ba & $\gamma$ & 356 keV & 10.54 y & 2 kBq \\ \hline
$^{137}$Cs & $\gamma$ & 662 keV & 30.2 y & 0.65 kBq \\ \hline
$^{22}$Na & $\gamma$ & $2\cdot 511$ keV + 1274 keV & 2.603 y & 11 kBq \\ \hline
\end{tabular}
\label{tbl:GammaSources}
\end{table}


\subsubsection{Neutron Sources}
\AmBe\ neutron sources of various activities and an \AmC\ neutron source have been deployed with CALIS to measure in-situ the \lsv's response to neutrons and the \tpc's response to nuclear recoils. The results in the \tpc\ can be compared to measurements performed in the \SCENE\ experiment \cite{Alexander:2013ke, Cao:2015ks}, which were used to determine the \tpc\ nuclear recoil energy scale and NR acceptance regions for the \dsf\ WIMP dark matter searches \cite{Agnes:2015gu, Agnes:2015_uar}.

\subsection{Calibration Campaigns Timeline and Stability}
The following calibration campaigns were performed between October 2014 and April 2016:
\begin{itemize}
\item The first extensive campaign involving all gamma sources and both a high and low activity \AmBe\ neutron source took place in October and November 2014 at LNGS. \AmBe\ source included a 2\,mm thick lead shielding to attenuate 59.6\,keV $\gamma$-rays from $^{241}$Am. The \tpc\ was filled with AAr with an inherent trigger rate of approx. 50 Hz from $^{39}$Ar. The \lsv's liquid scintillator consisted of PC, with $<0.1 \%$ TMB and 1.4 g/l PPO as wavelength shifter \cite{Agnes:2015qyz}.

\item A second campaign focusing on \lsv\ calibration using the low activity \AmBe\ source was performed in January and February 2015. Before this campaign, the \lsv\ was reconstituted with 5\% TMB, our nominal concentration \cite{Agnes:2015qyz}. Two deployments were performed at two different PPO concentrations (0.7 g/l and 1.4 g/l), allowing the study of the impact of the PPO concentration on alpha and gamma quenching. (1.4 g/l is our nominal PPO concentration, see also Fig.~\ref{fig:LSV:Calib}, right.)
\item A $^{22}$Na source was deployed next to the cryostat for TPC calibration in August 2015. This was the first gamma source calibration campaign after the deployment of UAr within \dsf.
\item An \AmC\ neutron source was deployed in December 2015, allowing an in-depth study of the detection efficiency of the prompt neutron thermalization signal, which is obfuscated in a majority of \AmBe\ decays by its neutron-correlated high energy gamma. \AmC\ source also utilized 2\,mm lead shielding to attenuate 59.6\,keV $\gamma$-rays.
\end{itemize}


\subsection{Calibration Results}
A few calibration results of the \tpc\ and \lsv\ are shown, illustrating the quality of acquired calibration data and the high level of agreement with their description in the G4DS MC package.

\subsubsection{\tpc\ Scintillation Energy of $^{57}$Co}
\begin{figure}[htbp]
\centering
\includegraphics[width=0.7\textwidth]{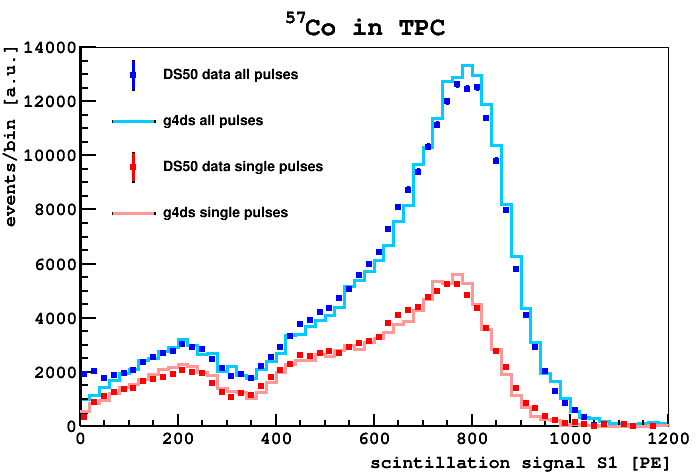}
\caption{Data-MC comparison for a $^{57}$Co source deployed next to the cryostat. In the red distribution a single-site interaction requirement is imposed as for dark matter events and for the blue distribution this constraint is removed \cite{DS50:G4DS:paper}. The data-MC discrepancy below 50 PE in the blue distribution (without single-scatter constraint) is attributed to fluctuations in the underlying background distribution surviving the background subtraction. It does not affect WIMP searches as the tuning of the MC-model is performed on single scatter data and matches the RoI of WIMPs above 50 PE. \label{fig:CalibData:Co57}}
 \end{figure}

Fig.~\ref{fig:CalibData:Co57} shows a data-MC comparison of the scintillation signal spectrum (S1) of a $^{57}$Co calibration source deployed next to the cryostat and at our central $z$-position. The S1 distribution is overlayed by an equivalent selection of G4DS MC simulation events.

\subsubsection{\tpc\ \FNinety\ Distribution from $^{241}$Am$^9$Be Neutron Data}\label{sec:CalibData:NR}
\begin{figure}[htbp]
\centering
\includegraphics[width=0.8\textwidth]{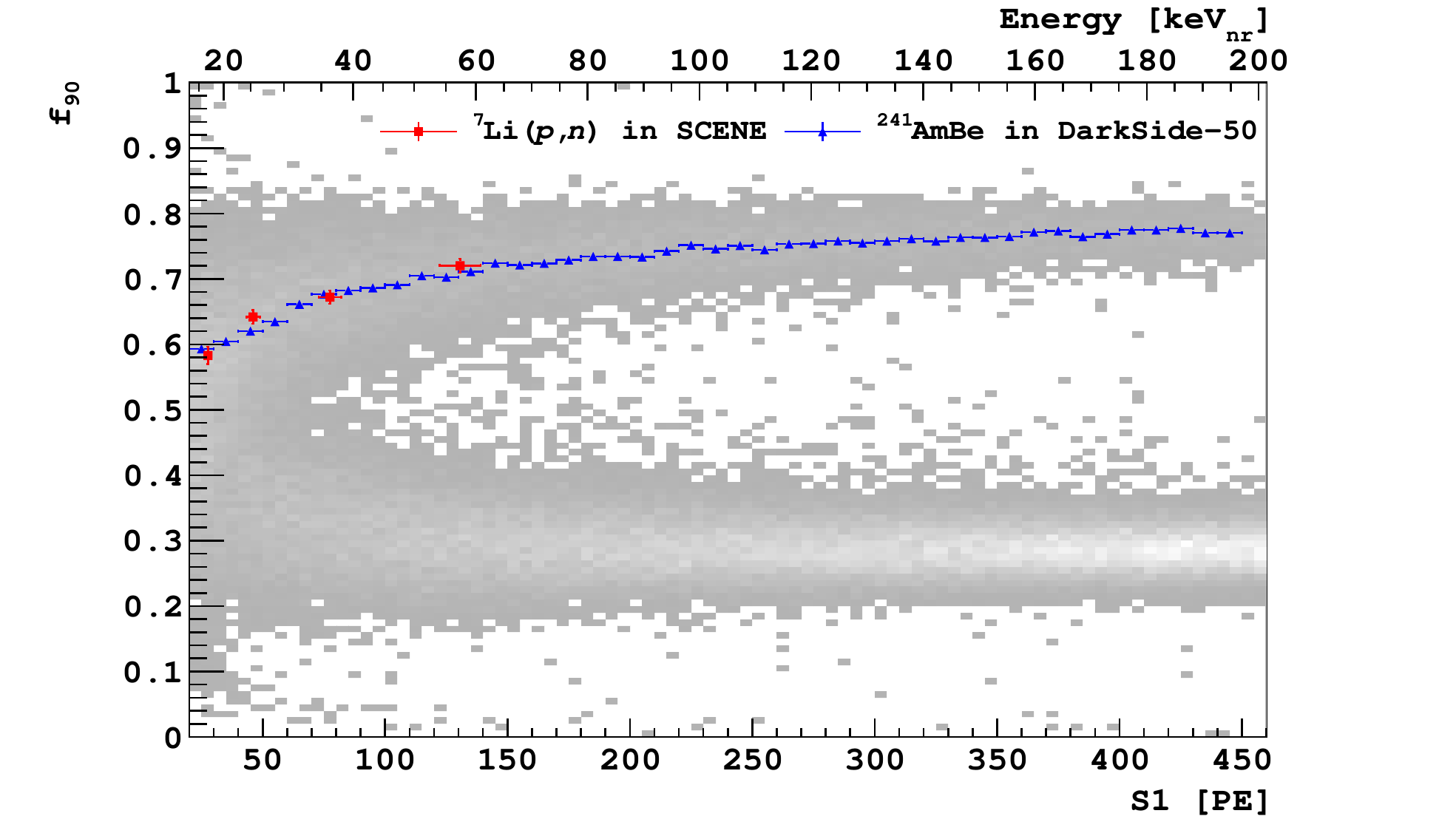}
\caption{Plot of \FNinety\ vs.~the scintillation signal S1 from a high rate {\it in situ} \AmBe\ neutron source calibration of \dsf\ in grey. The upper NR band from the \AmBe\ calibration and lower ER band from $\beta$-$\gamma$ backgrounds are visible. Overlaid are the \FNinety\ \NR\ medians vs.~\SOne\ from the \AmBe\ calibration (blue) and those from \SCENE\ measurements (red points) \cite{Alexander:2013ke, Cao:2015ks}. There is very good agreement between the two.  High source intensity and correlated neutrons and $\gamma$-ray emissions by the \AmBe\ source contribute events outside the nuclear recoil and electron recoil bands \cite{Agnes:2015_uar}.\label{fig:CalibData:F90}\label{fig:DSf-UArAmBeDMS}} 
\end{figure}

In our dark matter searches with AAr \cite{Agnes:2015gu} and UAr \cite{Agnes:2015_uar} a simple pulse shape disciminating parameter, \FNinety, has been used. \FNinety\ is defined as the fraction of the S1 signal that occurs in the first \timefno\ of the pulse relative to the S1 signal integral over \fixedintone. It is typically~$\sim$\num{0.3} for \bg-events and~$\sim$\num{0.7} for nuclear recoils. Fig.~\ref{fig:CalibData:F90} shows good agreement between \FNinety\ medians and the S1 spectra measured from \AmBe\ neutron data in the \dsf\ \tpc\ and those derived from \SCENE\ measurements \cite{Alexander:2013ke, Cao:2015ks}. This is a non-trivial and reassuring result given the broad neutron kinetic energy range of \AmBe\ source on the one hand, which is also high enough to produce correlated $\gamma$-ray emissions while passing through detector materials and on the other hand the \scene\ experiment optimized for the measurement of single-sited nuclear recoils, by exposing a small LAr \tpc\ to a low energy pulsed narrowband neutron beam of variable but fixed neutron kinetic energy.

\subsubsection{Source Position}\label{sec:SourcePosition}
\begin{figure}[htbp]
\centering
\includegraphics[width=0.48\textwidth]{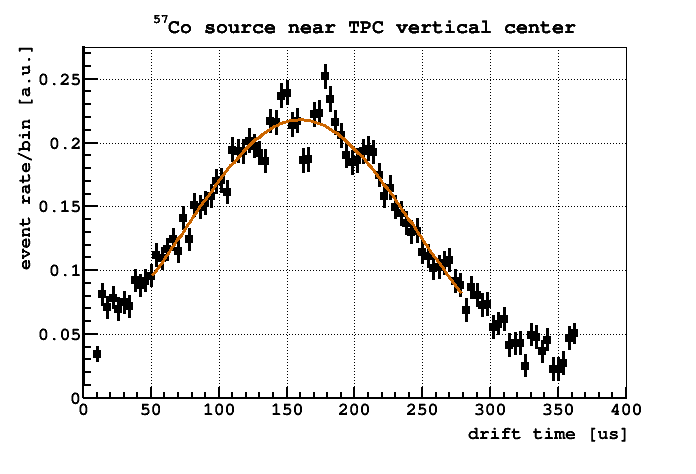}
\includegraphics[width=0.48\textwidth]{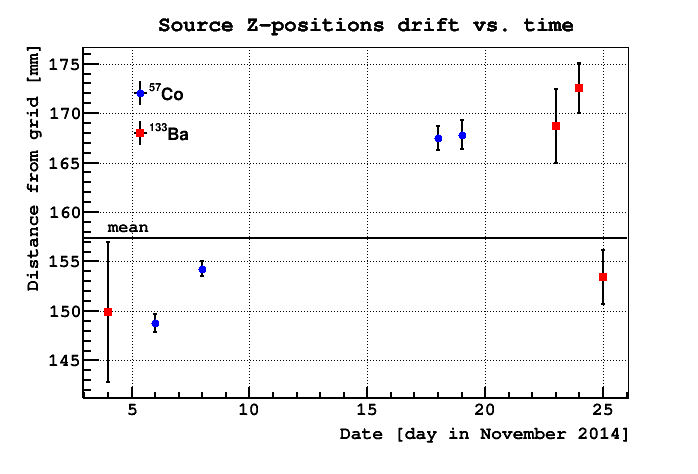}
\caption{\textit{Left:} A \tdrift\ distribution encoding the $z$-position of a $^{57}$Co source deployed next to the TPC vertical center. Single scatter events have been selected and a background distribution has been statistically subtracted. Fluctuations in the otherwise smooth \tdrift\ distribution are from copper field cage rings surrounding the TPC.
\textit{Right:} Shift of the source position relative to the TPC extraction grid as measured from the \tdrift\ distribution as a function of time when deployed to the same target position.
\label{fig:SourcePosition}} 
\end{figure}

Tests at LNGS established the deployment system's source positioning accuracy to be about $\pm$1 cm after a 7 meter journey into the \dsf\ \lsv.
During the first calibration campaign several runs have been taken with the source at its central position (as defined in Sec.~\ref{sec:CentralPosition}). 
Fitting the \tdrift\ distribution at that position for a sequence of runs, a systematic shift vs.~time has been observed (Fig.~\ref{fig:SourcePosition}, right). The source position has been on average 157.4 mm below the TPC extraction grid with an RMS of 10.1 mm, in cold conditions the TPC's active volume center is 172.7 mm below the grid. Following that observed systematic shift with time, the deployment procedures were revised to avoid such a time dependency in the future and to improve the deployment precision. Prior to moving the source to its target position, the deployment device is sent to its lowest position, where cables are fully unwound and any asymmetric build-up in the cables is released. It is worth mentioning that this source position imprecision does not induce significant uncertainties for calibration data analyses, as the \tdrift\ distribution can be measured in-situ on a per-run basis and offsets can be corrected for.

For the $xy$-position of the calibration source, distributions of the azimuthal angle in the $xy$-plane have been studied and a 139 degree mean was observed with a 1.2 deg RMS. (One degree corresponds to 6 mm at the outer cryostat, where the source is positioned.) However, an independent $xy$ reconstruction algorithm gave 142.5 degrees with an RMS of 0.8 deg, so that systematic uncertainties from the reconstruction algorithm dominate over the $xy$ positioning precision. 

\subsubsection{\lsv\ Integrity}
In dedicated analyses it has been shown that the calibration campaigns have not negatively affected the light yield so far or introduced additional radioactivity into the \lsv\ \cite{Agnes:2015qyz}.

\subsubsection{Liquid Scintillator Veto}\label{sec:LSV:gammasources}
\begin{figure}[htbp]
\centering
\includegraphics[width=0.44\textwidth]{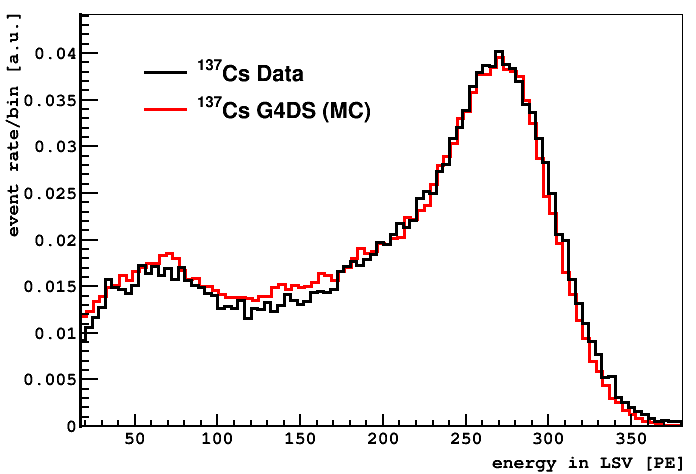}
\includegraphics[width=0.515\textwidth]{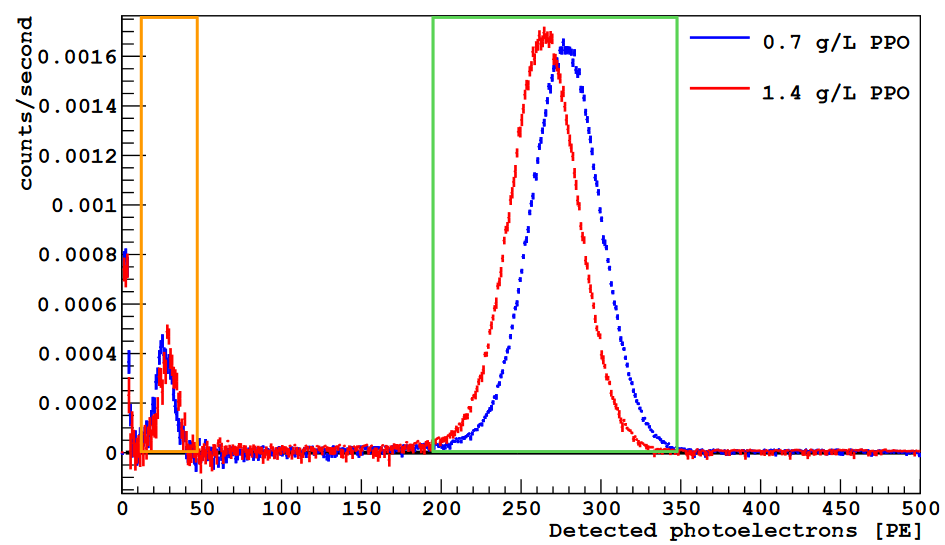}
\caption{\textit{Left:} A data-MC comparison of the $^{137}$Cs source LSV charge spectrum while the source was deployed next to the cryostat \cite{DS50:G4DS:paper}.
\textit{Right:} A clear detection signal of the neutron capture on $^{10}$B in the LSV is observed. The peak at $\approx$ 30 PE on the left (orange box) is due to a \enbortengroundalpha\ $\alpha$ + $^7$Li(g.s.). The peak on the right at $\approx$ 270 PE (green box) is from 93.6~\% of captures that lead to the $^7$Li excited state reaction, with the accompanying 478 keV $\gamma$-ray. The entries below 10 PE are due to PMT after-pulses. Data has been taken before and after varying the PPO wavelength shifter concentration in the scintillator with the \AmBe\ source rotated away from the cryostat to $\phi\,=\,90^{\circ}$. Though the additional PPO reduced the light yield a bit, it also reduced the quenching of the critical alpha, as expected. Thus at both PPO concentrations the de-excitation to ground state is clearly observed \cite{Agnes:2015qyz}.
\label{fig:LSV:Calib}} 
\end{figure}

In Fig.~\ref{fig:LSV:Calib} (left) a data-MC comparison of the LSV charge spectra from a $^{137}$Cs source deployed in the LSV next to the cryostat is shown \cite{DS50:G4DS:paper}.
In January and February 2015, the LSV scintillator reconstitution to 95\% PC with 5 \% TMB was completed and a second LSV calibration using an \AmBe\ neutron source was undertaken to further study various LSV neutron detection channels. With a borated scintillator, a critical aspect of the neutron detection efficiency is the capability to observe the \brbortenground\ capture branch leading to a \enbortengroundalpha\ $\alpha$ + $^7$Li(g.s.) without the accompanying 478 keV $\gamma$-ray. As shown in Fig.~\ref{fig:LSV:Calib} (right), the de-excitation channel is clearly observed at around 30 PE.

\section{Conclusions}\label{sec:Conclusions}\label{sec:Conclusion}
CALIS is a simple, affordable and effective source deployment system that has been successfully used to deploy sources in the LSV and next to the TPC and to conduct several successful calibration campaigns.



\section{Acknowledgements}\label{sec:Acknowledgements}
The DarkSide-50 Collaboration would like to thank LNGS laboratory and its staff for invaluable technical and logistical support. This report is based upon work supported by the US NSF (Grants PHY-0919363, PHY-1004072, PHY-1004054, PHY-1242585, PHY-1314483, PHY-1314507 and associated collaborative grants; grants PHY-1211308 and PHY-1455351), the Italian Istituto Nazionale di Fisica Nucleare (INFN), the US DOE (Contract Nos. DE-FG02-91ER40671 and DE-AC02-07CH11359), the Russian RSF (Grant No 16-12-10369), and the Polish NCN (Grant UMO-2014/15/B/ST2/02561). 
We thank the staff of the Fermilab Particle Physics, Scientific and Core Computing Divisions for their support. We acknowledge the financial support from the UnivEarthS Labex program of Sorbonne Paris Cit\'{e} (ANR-10-LABX-0023 and ANR-11-IDEX-0005-02) and from the S\~{a}o Paulo Research Foundation (FAPESP).



\bibliographystyle{veto-description} 
\bibliography{CALIS}


\end{document}